\documentclass[preprint2,twoside]{hwo}

\usepackage{graphicx}
\usepackage{float}
\usepackage{placeins}

\input{hwo.h}

\begin{document}

\title{\textbf{\LARGE Study of the atmospheric effects of energetic particle precipitations on giant planets with the Habitable World Observatory}}
\author {\textbf{\large J-Y. Chaufray$^{1}$,W. Dunn$^{2}$, L.N. Fletcher$^{3}$, L. Fossati$^{4}$, M. Galand$^{5}$, L. Gkouvelis$^{6}$,C.M. Jackmann$^{7}$, L. Lamy$^{8,9}$, L. Roth$^{10}$ }}
\affil{$^1$\small\it LATMOS-IPSL, UVSQ Paris Saclay, Sorbonne Université, CNRS, France \email{chaufray@latmos.ipsl.fr}}
\affil{$^2$\small\it MSSL, UCL, Dorking, UK}
\affil{$^3$\small\it SPA, University of Leicester, Leicester, UK}
\affil{$^4$\small\it IWF, OeAW, Graz, Austria}
\affil{$^5$\small\it Imperial College, London, UK}
\affil{$^6$\small\it LMU, University Observatory, Munich, Germany}
\affil{$^7$\small\it Astronomy and Astrophysics Section, Dublin Institute for Advanced Studies (DIAS), Dunsink Observatory, Dublin, Ireland}
\affil{$^8$\small\it LAM, Marseille, France}
\affil{$^9$\small\it LIRA, Meudon, France}
\affil{$^{10}$\small\it SPP, KTH, Stockholm, Sweden}


\author{\footnotesize{\bf Endorsed by:}
J-C. Bouret (LAM, Marseille, France), M. Davis (Southwest Research Institute, San Antonio, USA), B. Holler (STScI,Columbia,MD,USA), E. Lee (EisKosmos (CROASAEN), Inc), C. Neiner (LIRA, Meudon, France)
}



\begin{abstract}
  UV auroral emissions from giant planets are produced by extra-atmospheric energetic particles interacting with an atmosphere. They have been observed on Jupiter, Saturn and Uranus and should be present on Neptune. Even if the mechanisms are similar, each planet is unique due to its specific source of magnetospheric plasma and the structure and dynamics of its magnetosphere. How these precipitations modify atmospheric heating, dynamics and chemical balance at local and global atmospheric scale is still poorly known, especially on Uranus and Neptune, and critical to understanding the global atmosphere-magnetosphere system of giant planets and exoplanets. In this manucript we present how future observations by instruments, aboard \textit{the Habitable World Observatory} (HWO) will provide new information to better understand the origin and the atmospheric effects of these precipitations. A major interest is for the distant magnetospheres of Uranus and Neptune, never explored by an orbital spacecraft whose UV auroral emissions remains at (Uranus) or below (Neptune) the HST sensitivity. \textit{Pollux} is one such UV instrument concept, which will enable unprecedented high spectral resolution at fine spatial scale not previously seen and polarimetric observations of the planetary aurorae while \textit{LUMOS}, another UV instrument will image the full auroral regions with a good spectral resolution.	
 \textit{This article is an adaptation of a science case document developed for HWO's Solar System Steering Committee.}
  \\
  \\
\end{abstract}

\vspace{2cm}

\section{Science Goal}

\begin{figure*}[ht!]
    \centering
    \includegraphics[width=0.75\textwidth]{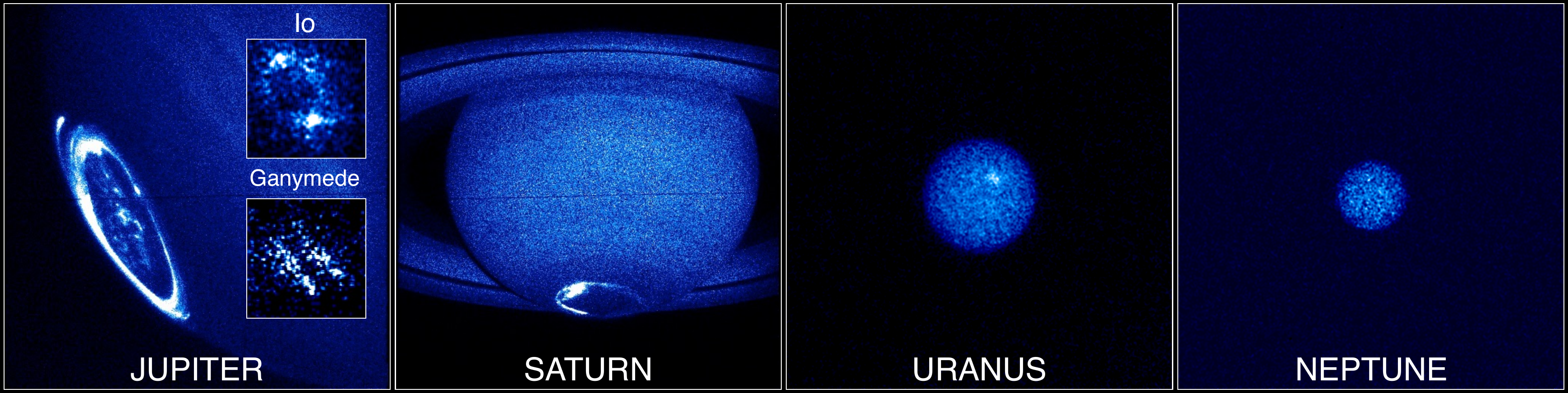}
    \caption{Examples of UV images of Jupiter (+ Io and Ganymede), Saturn, Uranus and Neptune obtained by HST showing some auroral emissions. Figure from \cite{Gomez_de_Castro_2022}}
    \label{fig:fig1}
\end{figure*}

UV auroral emissions from giant planets are produced by extra-atmospheric energetic particles interacting with an atmosphere. They have been observed on Jupiter, Saturn and Uranus and should be present on Neptune Fig~\ref{fig:fig1}. Even if the mechanisms are similar, each planet is unique due to its specific source of magnetospheric plasma and the structure of its magnetosphere.
The aurorae of Jupiter and Saturn are characterised by a combination of main oval emission, polar emission, diffuse equatorward emission, and discrete structures generated from satellites (e.g. footpoints and tails from Galilean moons at Jupiter and Enceladus at Saturn). These aurorae, observed from soft X-ray to radio, are produced by the precipitation of energetic magnetospheric particles (electrons and ions) \citep{Badman_2014, Grodent_2015, Lamy_2020}. These precipitations are a major source of energy into the atmosphere of the giant planets. They also mirror the magnetospheric processes and provide a means of probing the action of both internal and external drivers, from variable plasma loading and associated plasma injections to solar wind-magnetosphere coupling. The changes influenced by these drivers can be monitored both temporally (through brightenings/dimmings of specific regions) and spatially (e.g. through “auroral family” structures at Jupiter in UV and X-ray) \citep{Badman_2014, Grodent_2015, Grodent_2018, Lamy_2020, Dunn_2022}. However, how these precipitations modify atmospheric heating, dynamics and chemical balance at local and global atmospheric scale is still poorly known and critical to understanding the global atmosphere-magnetosphere system of giant planets and exoplanets.

The fundamental question we aim to address in this HWO science case is: What are the atmospheric effects of energetic particle precipitations on giant planets ?
This broad question can be split up into several questions, such as:

\begin{itemize}
    \item What is the energy of the precipitating particles ?
    \item How the atmospheric composition is modified by these precipitations ?
    \item How the atmospheric heating and its dynamics are modified by these precipitations ?
\end{itemize}

In particular, this Science Case is relevant to the following Key Science Questions and Discovery Areas of the Astro2020 Decadal Survey Report:

\begin{itemize}
    \item \textbf{E-Q2.} What are the properties of individual planets, and which processes lead to planetary diversity?
    \begin{itemize}
        \item \textbf{E-Q2c.} What fundamental planetary parameters and processes determine the complexity of planetary atmospheres?
        \item \textbf{E-Q2d.} How does a planet's interaction with its host star and planetary system influence its atmospheric properties over all time scales?
	\item \textbf{E-Q2e.} How do giant planets fit within a continuum of our understanding of all subsellar objects? 
    \end{itemize}
\end{itemize}
This Science Case is also relevant to the Future Needs mentionned in the Table E.2 of the Astro2020 Decadal Survey Report: 

\begin{itemize}
	\item UV space telescope: R $\>$ 1000 spectroscopy; high-contrast imaging of planets to detect UV absorbers	
	\item Ice giants: atmospheric and interior structure and composition.
\end{itemize}

\section{Science Objective}
\subsection{Determination of the local energy flux brought by energetic electrons in the auroral atmosphere by measurement of the electron energy}

\begin{figure*}[ht!]
    \centering
    \includegraphics[width=0.75\textwidth]{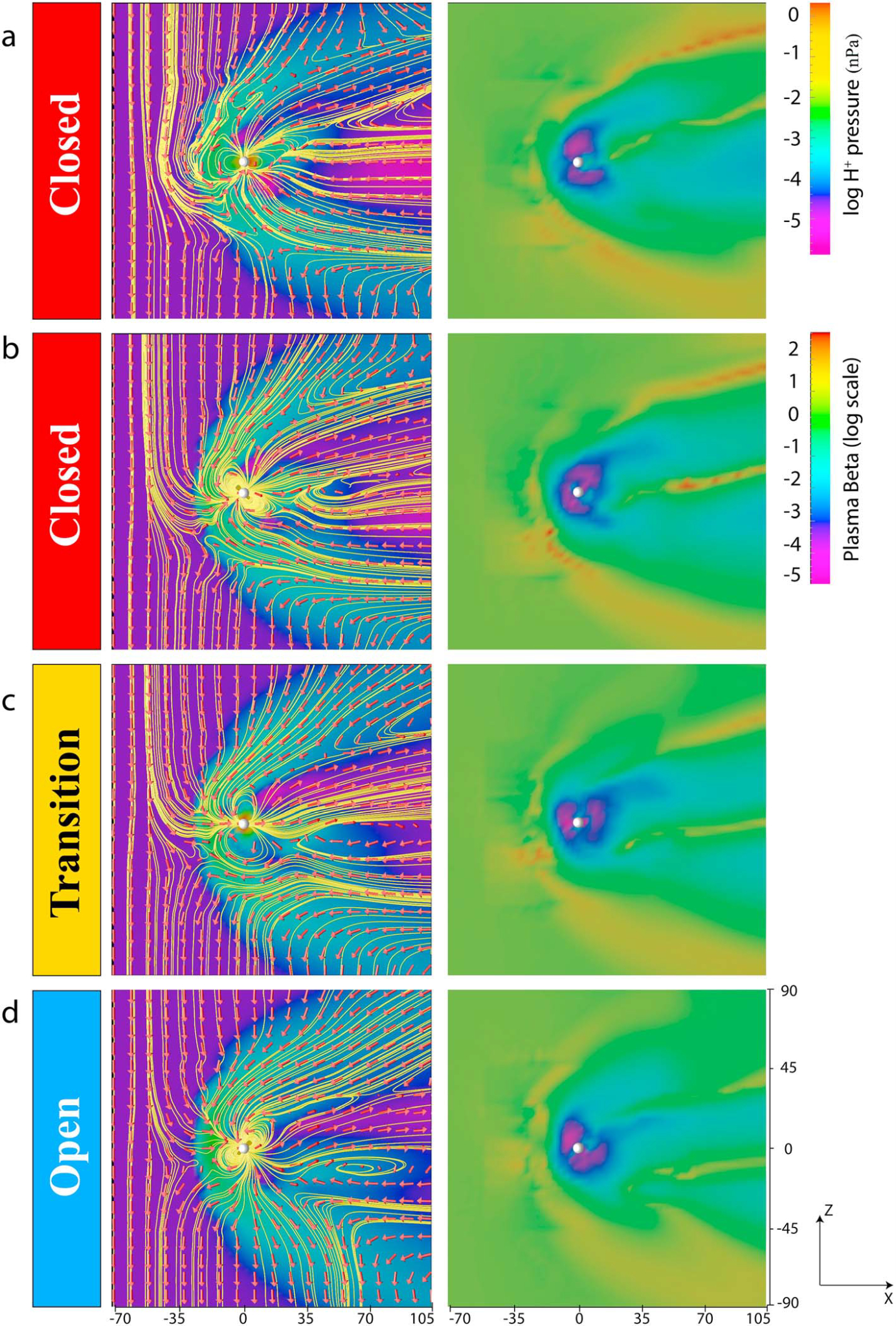}
    \caption{Logarithm of the H$^+$ pressure (left) and of the plasma beta (right) in the XZ plane at equinox. The magnetic field lines are represented in yellow and the red arrows represent the direction of the magnetic field. The global configuration of the magnetosphere is indicated by the label on the left. Figure from \cite{Cao_2017}}
    \label{fig:fig2}
\end{figure*}

The giant planets’ UV aurorae are mainly radiated from atmospheric H and H$_{2}$ species, collisionally-excited by accelerated charged particles precipitating along the auroral magnetic field lines. Aurorae thus directly probe complex interactions between the ionosphere, the magnetosphere, the moons, and the solar wind. The auroral emissions can be used to measure the energy and the flux of the precipitating particles. The energy of the electrons is critical not only to understand the details of the acceleration processes (for example the presence of field aligned currents) in the giant magnetospheres but also the local energy flux brought into the atmosphere.
The interaction between the solar wind and the oblique and fast rotating magnetic fields of the Ice Giants Uranus and Neptune is very different from the interaction with the quasi dipolar fields of Jupiter and Saturn and lead to rapidly evolving structures Fig~\ref{fig:fig2} (e.g. \citep{Cao_2017, Griton_2018, Pantellini_2020}.\\

For Uranus and Neptune, given the current low number of auroral observations, any new detections during seasons not observed yet would help to understand the basics of their magnetospheric processes. Observations of Jupiter, Saturn, Uranus and Neptune would be critical to understand the differences between the objects of the solar system and be used as large scale plasma laboratories to test mechanisms that could be applied outside the solar system (exoplanets, brown dwarfs in binaries)

\subsection{Determination of the atmospheric composition induced by the electron precipitations}

The composition of the upper atmosphere of the giant planets is dominated by H$_2$, H, He and several hydrocarbon species (CH$_4$, C$_2$H$_2$, ...). The local composition in some hydrocarbons (e.g. C$_2$H$_2$, C$_2$H$_4$) is modified in the jovian auroral regions due to the precipitation of energetic particles \citep{Kim_1985, Sinclair_2019}. 
The photochemistry needs to be supplemented by ion-neutral chemistry in auroral regions \citep{Guerlet_2015, Moses_2020} to explain enhanced species (e.g. benzene and heavier hydrocarbons on Saturn) in auroral regions. UV observations of different auroral regions can therefore provide insight into the interplay of aurorally-induced chemistry, planetary jets/banding and vertical diffusion in the upper atmosphere. Aerosols in Jupiter’s stratosphere form polar hoods, possibly composed of fractal aggregates formed by condensation of hydrocarbons at low temperatures \cite{Zhang_2015} that could be driven by magnetospheric – ionospheric interactions \cite{Tsubota_2025}. In addition, UV multispectral image observations (including the most recent from NASA’s JUNO spacecraft) have been used in the past to infer the electron energy ﬂux and their characteristic energy and the  vertical distributions of H$_3$$^+$ and hydrocarbon ions from infrared observations \citep{Gustin_2017, Gerard_2020, Gerard_2023}. Finally, other species like NH$_3$ ($\ge$ 160 nm) or PH$_3$ (160-180 nm) in the giant planets’ stratospheres have spectral signatures in UV range and could be used to determine the composition of the atmosphere in the auroral regions.

\subsection{Determination of the heating induced by the energetic electrons and its consequences on the local/global thermal balance and dynamics of the atmosphere}

The precipitation of energetic electrons in the upper atmosphere of the giant planets is also a major source of local heating of the upper atmosphere and stratosphere, as deep as millibar pressures on Jupiter \cite{Sinclair_2018}. This energy input has been suggested to be the major source of heating able to explain the high temperature in the thermosphere of giant planets \cite{Majeed_2004}. The auroral observations suggest that the upper atmosphere of Jupiter in the auroral regions is much hotter than in the non-auroral regions \cite{Trafton_1994}. The auroral heat can be redistributed to lower latitudes  by intense thermospheric winds \cite{Hue_2024}. 
The full study of the auroral regions would:
\begin{itemize}
	\item \textit{Determine the local energy and flux of the precipitating energetic electrons}
	\item \textit{Determine the local atmospheric composition}
	\item \textit{Determine the local temperature and winds velocity}
\end{itemize}
Such study could be done at different auroral regions of the four giant planets and at non-auroral regions for reference to enhance the local atmospheric effects of the precipitating electrons.
A possible expansion of this study could also include observations of the low-latitude ionosphere to better know the global context or/and look for structures in the H$_2$ emissions. The spectro-imager \textit{LUMOS} could also be used to partly do the mentionned science on a much larger area including auroral and non-auroral regions.

\section{Physical Parameters}

\subsection{The determination of the energy distribution of the electrons will require measurement of the UV spectra in auroral regions with high spectral resolution and polarimetric observations}

The UV spectra of the giant planets between 110 – 200 nm show several emission lines of H$_2$ excited by energetic electron impact associated with the Lyman and Werner bands \citep{Liu_1996, Benmahi_2024}. The relative emissions of the lines strongly depend on the energy of the precipitating electrons. Moreover, the color ratio CR of the ratio between the brightness I(155-162 nm) divided by the brightness I(123-132 nm) can be used to derive independent information on the energy of the electrons. Indeed, the energetic electrons can penetrate deeply in the atmosphere where the absorption of H$_2$ emissions at $\lambda$ $\le$ 140 nm by CH$_{4}$ is important and the CR of the auroral emissions is high. The less energetic electrons excite molecules at higher altitudes where CH$_{4}$ absorption is negligible and the CR is low \citep{Barthelemy_2014, Gustin_2017, Gerard_2019}. The brightness ratio of H-Lyman-$\alpha$ to H$_{2}$ bands can also be used to directly probe the low energy electrons e.g. \cite{Tao_2014}. Finally, laboratory experiments and theory indicate that electron-impact excitation of atomic hydrogen will polarize the Lyman-$\alpha$ emission with a degree of polarization dependent on the electron energy \cite{James_1998}.

\subsection{The determination of the hydrocarbon composition of the atmosphere can be studied from the presence of absorption lines, the D/H ratio from the spectral separation of the D and H emission lines, and polar haze abundance from polarimetric and absorption observations between 110 - 500 nm}

\begin{figure*}[ht!]
    \centering
    \includegraphics[width=0.75\textwidth]{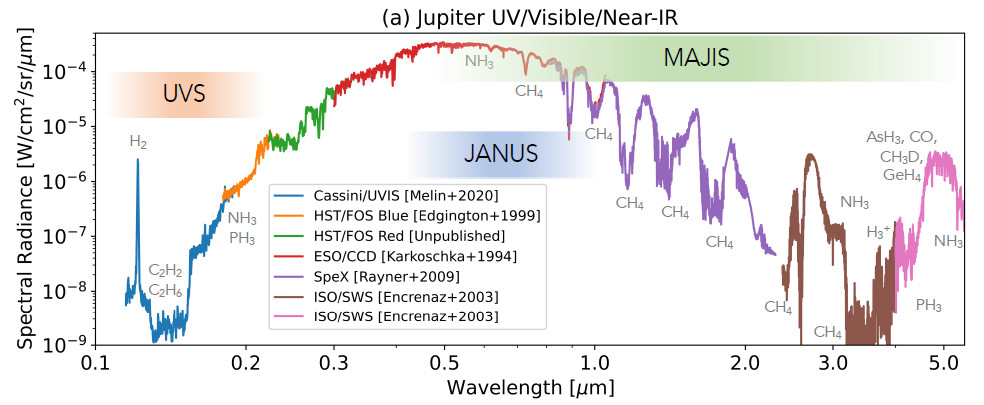}
    \caption{Spectral Radiance of Jupiter from 120 nm to 5000 nm. The absorption bands due to C$_2$H$_2$, C$_2$H$_6$ (in UV) and CH$_4$ (visible and infrared) are indicated. Note also some absorption bands due to NH$_3$ and PH$_3$ in the UV range ($\le$ 200 nm). Figure from \cite{Fletcher_2023}}
    \label{fig:fig3}
\end{figure*}

The spectrum between 100 and 200 nm also presents several absorption lines due to different species (CH$_{4}$, C$_{2}$H$_{2}$,...) Fig.~\ref{fig:fig3}. The spectral position of these absorption lines can be used to derive the composition of the local atmosphere and the magnitude of the absorption used to derive the abundance of the different species. The D and H Lyman-$\alpha$ lines are separated by $\sim$ 0.3 A, and therefore a spectrometer with a spectral resolution larger than 50,000 like POLLUX can separate both lines and be used to measure the D/H from these bright emission lines, if the sensitivity is sufficient to measure the D Lyman-$\alpha$ emission. UV polarimetric observations have not been used frequently to study the upper atmospheres of the giant planets \citep{Hord_1979, Pryor_1991, Pryor_1992} but, while Rayleigh scattering is important in the UV \cite{West_1981}, aerosol scattering is important in the visible \cite{West_1980} and be a major source of polarization of the reflected solar light.

\subsection{The determination of the temperature and winds velocity can be derived from the line shape of the H$_{2}$ rotational lines or from the Lyman-$\alpha$ spectral profile}

The H$_2$ spectrum is dominated by the Lyman and Werner band system composed of several rotational lines observable at high spectral resolution. From these lines, the rotational temperature, representative of the local temperature, just above the homopause (~10$^{-5}$ to 10$^{-8}$ bar), can be derived \citep{Clarke_1994, Liu_1996} in complement to hydrocarbon infrared measurements (~10$^{-4}$ to 10$^{-6}$ bar) or H$_3$$^{+}$ infrared emission (~10$^{-6}$ to 10$^{-10}$ bar) \cite{Grodent_2001}. Auroral observations at several spots (along oval, at satellite spots,..) on the Jovian disk could be used to study possible local enhancement of the temperature and constrain the heating source associated with energetic electron precipitations. Finally, an instrument with a very high spectral resolution $R \ge$ 20,000 able to resolve the hydrogen Lyman-$\alpha$ line (few 10s mA at 1216 A) could detect asymmetries in the line spectral profile giving information on the winds in the upper atmosphere \citep{Prange_1997, Chaufray_2010}.

\subsection{Observational State of the Art and Science Targets}

The expected instrumental performances are summarized in Table \ref{tab:performance}

\begin{table*}[htpb]
    \centering
    \caption[Performance Goals]{}
    \label{tab:performance}
	\begin{tabular}{p{3.0cm}p{3.0cm}p{3.0cm}p{3.0cm}p{3.0cm}}
        \noalign{\smallskip}
        \hline
        \noalign{\smallskip}
        {Physical Parameter} & {State of the Art} & {Incremental Progress} & {Substantial Progress} & {Major Progress} \\
        \noalign{\smallskip}
        \hline
        \noalign{\smallskip}
	    Electron energy & Juno, Cassini, HST... & Color ratio & Observations of numerous individual H$_2$ rotational lines & Measurements of the spectral linewidth of the H$_2$ rotational lines \\
        Atmospheric Composition & Juno, Cassini, HST, JWST...  & Reflectance between 160 - 500 nm with a spectral resolution of 0.2 nm & Observations of the individual absorption lines of hydrocarbon species & Observations of both brightness and polarization at spectral resolution of ~ 0.002 nm \\
		Temperature (~10$^{-5}$ to 10$^{-8}$ bar) & Juno, Cassini, JWST, ESO-VLT,...& Reflectance between 160 - 500 nm with a spectral resolution of 0.2 nm & Observations of numerous individual H$_2$ rotational lines & Measurements of the spectral linewidth of the H$_2$ rotational lines \\ 
        \noalign{\smallskip}
        \hline
	\end{tabular}
\end{table*}

\section{Description of Observations}

\subsection{Measurement of H$_2$ and H emission lines between 110 – 200 nm}

The FUV (110 – 120 nm) and MUV (118 – 236 nm) arms of POLLUX encompass a spectral range including the H Lyman-$\alpha$ line (121.6 nm) and numerous H$_{2}$ rotational lines of the Lyman and Werner system \cite{Liu_1996} with a very high spectral resolution ($\delta \lambda$ $\sim$ 10 mA in FUV and $\sim$ 20 mA in MUV), the CR as the H/H$_2$ ratio will be used to derive the energy of the precipitating electron on the different auroral regions of the giant planets. The individual lines’ strength would also provide strong constraints on the energy of the precipitating electrons. The FoV of POLLUX ($\sim$ 0.04’’) is much lower than the angular diameter of all the giant planets, and the auroral regions should be targeted accurately.
To point a specific auroral region on a planet, the strategy is to first point a smaller target (for example one satellite) and then apply a calculated small shift to point the expected auroral region.
While on Jupiter and Saturn the auroral locations are well known and the flux is large $\sim$ 10$^{12}$ erg/cm2/s/A, the observation could be more difficult for Uranus and Neptune where the auroral regions are lesser known.  On Uranus, the recent accurate rotation period derived by \cite{Lamy_2025} from auroral detections, leads up to a new longitude model valid over decades that could be used to locate the auroral regions.

\subsection{Measurement of the spectral absorption lines between 160 – 200 nm}

\begin{figure*}[h!]
    \centering
    \includegraphics[width=0.7\textwidth]{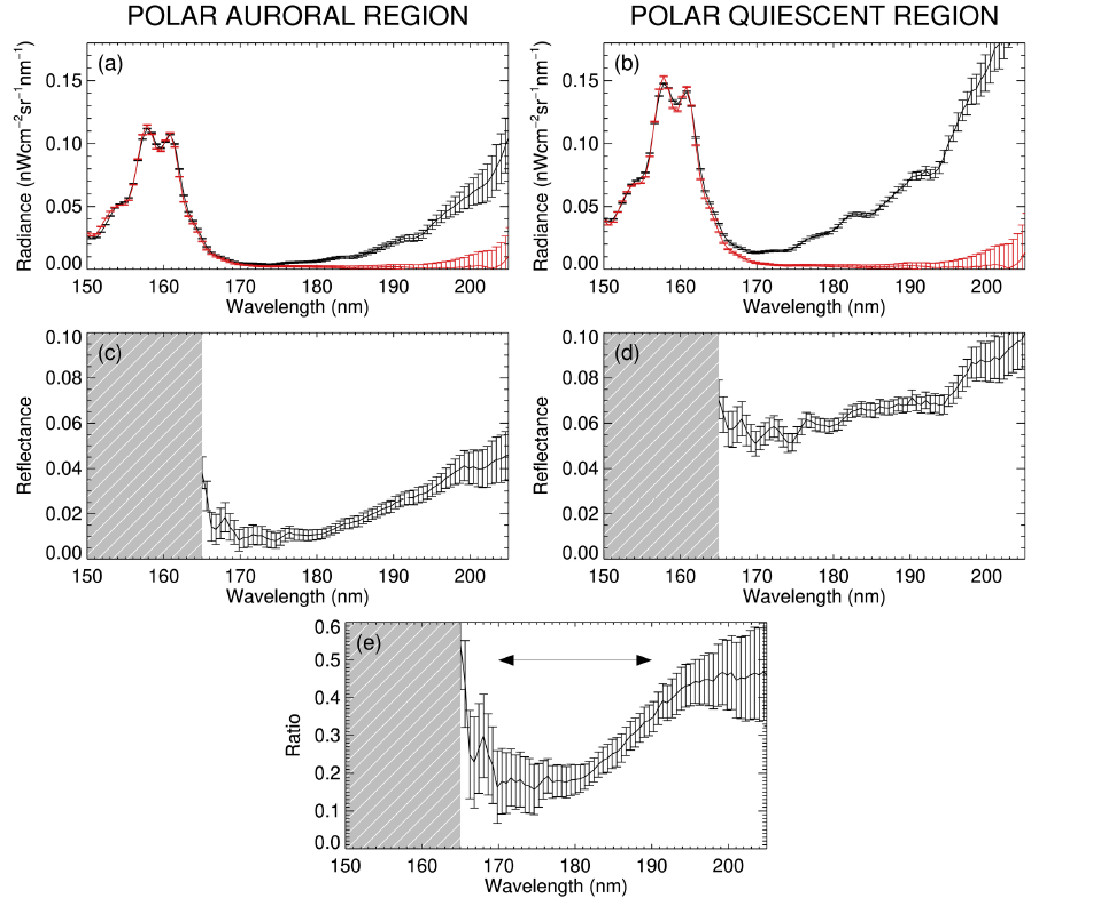}
    \caption{Comparison between the UV reflectance on polar auroral region and polar quiescent region showing a larger absorption (larger concentration of hydrocarbons) in the auroral region \cite{Giles_2023}}
    \label{fig:fig4}
\end{figure*}

The spectral range between 160 and 200 nm contains several absorption lines from C$_2$H$_2$ and C$_2$H$_6$ that can be measured with POLLUX. The most recent derivation of the hydrocarbon contents from the UV band has been done by JUNO with a spectral resolution $\delta \lambda$ $\sim$ 2 – 20 A, not high enough to distinguish the absorption lines from C$_2$H$_2$ and C$_2$H$_6$ (Fig.~\ref{fig:fig4}).

\subsection{Measurement of the spectral linewidth of the emission lines between 120 – 200 nm}

The temperature of Jupiter's upper atmosphere varies between ~100 K to 1100 K from the stratosphere to the exobase. At a temperature T = 300K, the H$_{2}$ linewidth is $\sim$ 8 mA. This value is two times lower than the spectral resolution of POLLUX near 150 nm and then a direct derivation of the temperature from the linewidth could be attempted if the instrumental line spread function is well known. Another possible method is to derive the rotational temperature from the relative magnitude of the H$_{2}$ emission lines using synthetic spectra depending on electron energy and temperature that best fits the observed spectra similar to the approach used by \cite{Liu_1996} to analyze the HST/GHRS spectra at spectral resolution of 560 mA, a resolution 28 times lower than Pollux. A spectral resolution of 70 mA has been shown to be sufficient to observe the self-reversed asymmetric Lyman-$\alpha$ spectral line and to derive the hydrogen wind velocity \citep{Prange_1997, Chaufray_2010}. A spectral resolution of 70 mA or better ($R \ge$ 20,000) is needed to derive the wind velocity. Such measurements are very rare and any new observations will provide important information on the wind magnitude in the thermosphere of Jupiter.

\subsection{Observational Requirements}

The observation requirements are summarized in Table \ref{tab:obsreq}

\begin{table*}[t!]
\centering
\caption[Observation Requirements]{}
\label{tab:obsreq}
\begin{tabular}{ccccc}
        \noalign{\smallskip}
        \hline
        \noalign{\smallskip}
        {Observation} & {State of the Art} & {Incremental Progress} & {Substantial Progress} & {Major Progress} \\
        \noalign{\smallskip}
        \hline
        \noalign{\smallskip}
        Spectral range & 100 - 200 nm & 100 - 200 nm & 100 - 500 nm & 100 - 1000 nm \\
        Spectral Resolution & HST/GHRS $\sim$ 70 mA & 70 mA & 50 mA & 20 mA \\
	Sensitivity & JUNO, Cassini, HST & 10$^{-12}$ erg/cm2/s/A & 10$^{-13}$ erg/cm2/s/A & 10$^{-14}$ erg/cm2/s/A \\
        \noalign{\smallskip}
        \hline
\end{tabular}
\end{table*}

{\bf Acknowledgements.}

\bibliography{author.bib}

\end{document}